# End-to-End Automatic Sleep Stage Classification Using Spectral-Temporal Sleep Features

Hyeong-Jin Kim, Minji Lee, and Seong-Whan Lee, *Fellow, IEEE*

*Abstract*— Sleep disorder is one of many neurological diseases that can affect greatly the quality of daily life. It is very burdensome to manually classify the sleep stages to detect sleep disorders. Therefore, the automatic sleep stage classification techniques are needed. However, the previous automatic sleep scoring methods using raw signals are still low classification performance. In this study, we proposed an end-to-end automatic sleep staging framework based on optimal spectral-temporal sleep features using a sleep-edf dataset. The input data were modified using a bandpass filter and then applied to a convolutional neural network model. For five sleep stage classification, the classification performance 85.6% and 91.1% using the raw input data and the proposed input, respectively. This result also shows the highest performance compared to conventional studies using the same dataset. The proposed framework has shown high performance by using optimal features associated with each sleep stage, which may help to find new features in the automatic sleep stage method.

*Clinical Relevance*— The proposed framework would help to diagnose sleep disorders such as insomnia by improving sleep stage classification performance.

## I. INTRODUCTION

Sleep is an important factor that directly affects our health and quality of life [1]. However, sleep disorders are widespread in most people and can cause serious health problems that affect the quality of life [2]. Some of these sleep disorders should be diagnosed using advanced techniques [2, 3]. Polysomnography (PSG) is one of the experiments with advanced technology that requires multi-modal bio-signals such as electroencephalogram (EEG), electrooculogram (EOG), electrocardiogram (ECG), and electromyogram (EMG) [4, 5]. In a conventional process, the sleep experts score and grade the sleep stages in manually [6]. However, it is very inefficient and costly to inspect the PSG signals and classify the sleep stages manually.

To alleviate this, automatic sleep stage classification using deep learning has recently been widely attempted [7-11]. Indeed, the excellence of deep learning has already been demonstrated in many studies using EEG signals [12, 13]. Supratak et el. [7] proposed DeepSleepNet, which uses raw single-channel EEG. It was the first attempt in that field to use deep learning. Even so, the study suggested quite a complex network, including the use of two-stream convolutional neural networks (CNN) and bidirectional long short-term memory (LSTM). Phan et al. [8, 9] also proposed a novel CNN framework. They exploited one EEG channel and one EOG channel with a CNN architecture for classification. Additionally, they added a special layer, named multi-task softmax layer which is suitable for joint classification and prediction. It works with the multi-task loss function which penalizes for both misclassification and misprediction on a training example. In the most recent study, Patanaik et al. [10] and Yildirim et al. [11] proposed a deep learning model for automated sleep stages classification. They also used CNN for solving this issue. Unlike previous studies, they stacked lots of layers to extract the most effective features from the input data. It works more successfully than the other methods in terms of overall accuracy. However, previous methods still have low sleep stage classification performance for practical use.

In general, there are several criteria for classifying sleep stages. Two most dominant criteria are R&K standard, which was proposed by Rechtschaffen and Kales [14], and AASM which was developed by the American Academy of Sleep Medicine [15]. We followed the more up-to-date standard of this, AASM for classifying sleep stages. In AASM criterion divided sleep into five stages: wake (W), three non-rapid eye movement (NREM) stages (N1-N3), and rapid eye movement (REM) [1]. During W stage, alpha activity (8-12 Hz) becomes prominent, particularly in the occipital region [16]. N1 is a transitional stage between W and N2. This stage is characterized by the loss of alpha activity and the appearance of theta activity (4-8 Hz) [1]. N2 is the stage where the actual sleep begins, this stage produces a unique frequency waveform called the sleep spindle (12-15 Hz) [17]. N3 is considered as deep sleep because the function of the brain is significantly reduced. In this stage, mainly the delta wave (0.5-4 Hz) with the strongest amplitude appears [18]. Lastly, the low-voltage and fast activity reappear in the REM sleep, especially increased power in the theta waves [19]. Although there are clear frequency characteristics for each sleep stage, previous studies have not used these features so far.

In this paper, we propose an automatic sleep stage classification framework using spectral-temporal sleep features. We hypothesized that using the optimized temporal-spectral features can improve the performance for the automatic sleep stage classification higher. This study would help determine optimal features associated with sleep for automatic sleep stage classification.

*Research supported by Institute for Information & Communications Technology Promotion (IITP) grant funded by the Korea government (No. 2017-0-00451; Development of BCI based Brain and Cognitive Computing Technology for Recognizing User`s Intention using Deep Learning).

H.-J. Kim and M. Lee are with the Department of Brain and Cognitive Engineering, Korea University, 145, Anam-ro, Seongbuk-gu, Seoul 02841, Republic of Korea (e-mail: kme0115@korea.ac.kr (H.-J. Kim), minjilee@korea.ac.kr (M. Lee)).
S.-W. Lee is with the Department of Artificial Intelligence, Korea University, 145, Anam-ro, Seongbuk-gu, Seoul 02841, Republic of Korea (corresponding author: sw.lee@korea.ac.kr).

## II. MATERIALS AND METHODS

### A. Dataset

The sleep-edf (expanded) dataset [20] was used for our study. The 145 out of 153 PSG data were included because of excluding 8 which have problems to load. It was composed of 77 subjects who are between 25 and 101 years (40 females, average age 58.7). These PSG recordings include two bipolar EEG channels (Fpz-Cz and Pz-Oz), one horizontal EOG channel, and one submental chin EMG channel. The EEG and EOG signals were each sampled at 100 Hz, and EMG signals were sampled at 1 Hz. Also, each 30 seconds fragment was scored based on the R&K manual. Also, we used one EEG channel (Fpz-Cz) and one EOG channel (horizontal). Fig. 1 denotes that its hypnogram of the sample PSG signals of SC4071EC.

### B. Proposed Features

A bandpass filter was applied to preprocess raw input data [21]. The input of the model consisted of a matrix with six rows, each row is sequentially raw EEG signal (Fpz-Cz), bandpass filtered EEG signals according to each band (delta: 0.5-4 Hz, theta: 4-8 Hz, alpha: 8-12 Hz, and sleep spindle: 12-15 Hz), and raw EOG signal (horizontal) (Fig. 2). This input data was constructed in that the specific frequency appears prominently depends on the sleep stage [19]. We compared in two cases: the control and the experimental groups. In the control group, we used the EEG and EOG signals applied to the notch filter at 50 Hz to remove power-line noise. On the other hand, the experimental group was adjusted by applying the proposed method as the input data.

### C. CNN Architecture

We used a CNN that is composed of four convolution layers and two max-pooling layers (Table I). This architecture was inspired by the representation learning of DeepSleepNet [7]. At the first layer, we applied a kernel with a size of 200 because the dataset was recorded 100 Hz and the lowest bandpass that we considered is 0.5 Hz. Then, the smaller filter size was applied at the following layer to capture well the temporal information. By increasing the kernel size as the layer deepens, the model can learn different frequency features according to the sleep stage [7]. However, because the size of the input data in control and experimental groups are different, the used model was inevitably different. We kept other parameters and adjusted only the height of the filter size to minimize the variation of the model. The values in parentheses at the Table I are the parameters of the model for the control

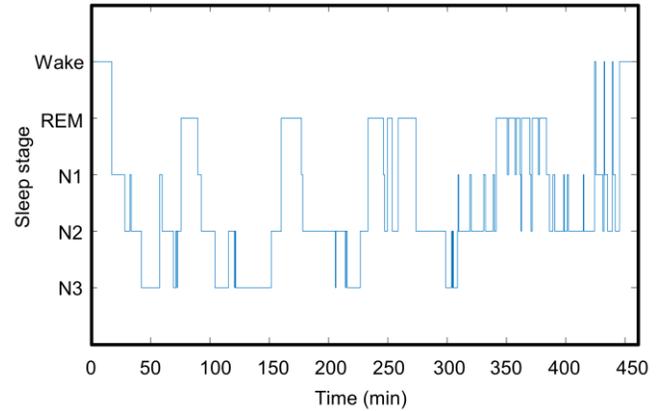

Figure 1. The hypnogram of sample PSG signals records obtained from the sleep-edf (expanded) database. This data of SC4071EC were recorded for 7.7 hours and the ground truth label was changed to meet the AASM standard.

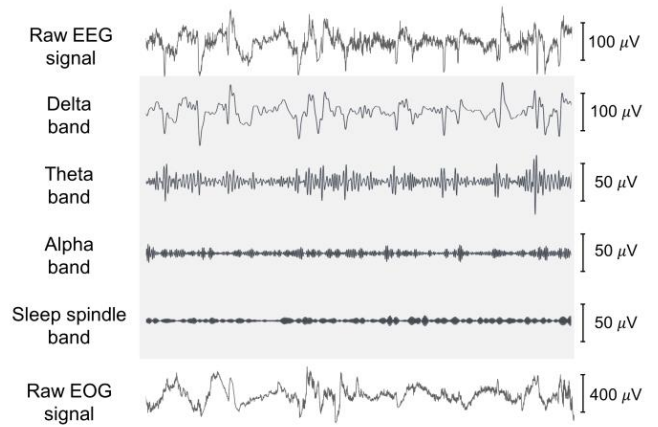

Figure 2. The preprocessed input data. The first and last rows represent the raw EEG (Fpz-Cz) signal and the raw EOG (horizontal) signal, respectively. The gray area represents the bandpass filtered data in the delta (0.5-4 Hz), theta (4-8 Hz), alpha (8-12 Hz), and sleep spindle (12-15 Hz), respectively.

group; those are the model for the experimental group.

For all experimental results, every dataset was repeatedly trained using 50 epochs. We used cross-entropy as a loss function. Also, the Adam optimizer was selected for boosting the learning process of the proposed model. For hyperparameters, we used a batch size of 10, a learning rate of 0.00001 and weight decay by 0.003. For evaluating the proposed method, we applied 20-fold cross-validation. The $\kappa$-value and classification accuracy were used for the performance measurement criteria [22, 23].

TABLE I. DETAILS OF LAYERS AND PARAMETERS USED IN THE PROPOSED CNN MODEL

| Number | Layer type | Number of filters | Kernel size | Stride | Activation function | Output size |
|---|---|---|---|---|---|---|
| 1 | Convolution 1 | 10 | $200 \times 1$ | $1 \times 1$ | ReLU[a] | $10 \times 2801 \times 6$ ($10 \times 2801 \times 2$) |
| 2 | Max-pooling 1 | 10 | - | $2 \times 1$ | - | $10 \times 1400 \times 6$ ($10 \times 1400 \times 2$) |
| 3 | Convolution 2 | 20 | $32 \times 2$ ($32 \times 1$) | $1 \times 1$ | ReLU | $20 \times 1369 \times 5$ ($20 \times 1369 \times 2$) |
| 4 | Convolution 3 | 30 | $128 \times 2$ ($128 \times 1$) | $1 \times 1$ | ReLU | $30 \times 1242 \times 4$ ($30 \times 1242 \times 2$) |
| 5 | Convolution 4 | 40 | $512 \times 4$ ($512 \times 2$) | $1 \times 1$ | ReLU | $40 \times 731 \times 1$ |
| 6 | Max-pooling 2 | 40 | - | $2 \times 1$ | - | $40 \times 365 \times 1$ |
| 7 | Fully-connected 1 | 1 | - | - | ReLU | $100 \times 1$ |
| 8 | Fully-connected 2 | 1 | - | - | Softmax | $5 \times 1$ |

a. Rectified linear unit.

$$\kappa \equiv \frac{P_0 - P_e}{1 - P_e} = 1 - \frac{1 - P_0}{1 - P_e}. \quad (1)$$

where $P_0$ and $P_e$ denote the accuracy and the probability of chance rate, respectively.

*D. Spectral Power Analysis*

We additionally analyzed the power spectral density (PSD) in each stage to investigate what the CNN model is learning [5, 17]. It is widely used to describe how the power of the signal or time series is distributed over frequency [24-25]. The PSD was calculated in the delta, theta, alpha, sleep spindle bands like the proposed feature.

## III. RESULTS

*A. Classification Performance for Five Sleep Stages*

The results are presented in the form of a normalized confusion matrix in Fig. 3. According to the result, overall accuracy and $\kappa$-value were 85.6%, 0.82 for the control group and 91.1%, 0.889 for the experimental group, respectively. Both groups showed high performance in W and N3, while relatively low performance in REM, N1, and N2. The lowest classification performance was N1 with 58.5% and 71.5% in the control and experimental groups, respectively. Eventually, the performance differences between the two groups represented in N1 and N2. The N1 was mainly misclassified by REM, and N2 was misclassified primarily by N1.

*B. Performance Comparison with Other Methods*

Table II indicates the five-class classification performance and the accuracy of each class over the previous studies using the sleep-edf (expanded) dataset. The proposed method shows the highest accuracy compared to the related works for N1, N3 and overall accuracy (71.5%, 95.8% and 91.1%, respectively). Also, the $\kappa$-value is 0.889, indicating the highest value than other methods. However, it represented relatively low performance than other methods for REM and N2 stages.

*C. Spectral Power in Five Sleep Stages*

We investigated the PSD in each sleep stage (Fig. 4). As the sleep progresses from the W stage to the N2 stage, the power of the alpha band is weakened, and the power of the sleep spindle band is strengthened. In the N3 stage, delta power was increased, but other powers were decreased. Lastly, in the REM stage, theta and alpha powers became increased. In summary, each sleep stage has mainly spectral power.

## IV. DISCUSSION AND CONCLUSION

In this study, we proposed a new end-to-end automatic sleep stage classification method using optimal spectral-temporal features. As a result, the classification performance using the proposed model was much higher than the conventional performance, especially in the N1 stage.

We considered two reasons for higher performance. First, we used optimal input data for sleep classification. In each sleep stage, the prominent frequency bands clearly appear [19, 26]. By using the optimal feature associated with sleep, the proposed model seems to learn the spectral characteristics in each sleep stage. Second, an appropriate CNN model was applied to deal with the modified input data. The used model was designed to capture the time-series features of the input data at the shallow layers and the frequency-domain features at the deeper layers [7].

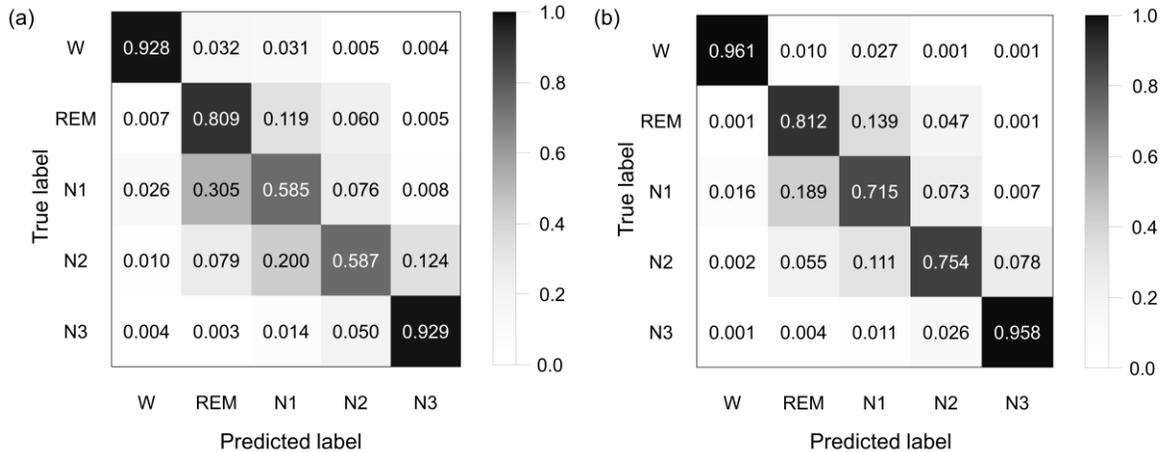

Figure 3. The normalized confusion matrix of (a) the control group and (b) the experimental group. Each number in the cell is the accuracy according to each class. The columns denote predicted labels and the rows are true labels. (W: wake, REM: rapid eye movement, N1: Non-REM 1, N2: Non-REM 2, N3: Non-REM 3)

TABLE II. COMPARISON BETWEEN THE PROPOSED METHOD AND OTHER METHODS USING THE SLEEP-EDF DATASET

| Study | Number of Channels | Method | $\kappa$-value | Accuracy (%) | | | | | |
|---|---|---|---|---|---|---|---|---|---|
| | | | | Overall | W | REM | N1 | N2 | N3 |
| Supratak et al. [6] | 1 EEG[a] | CNN[c] + bi-LSTM[d] | 0.775 | 82.0 | 83.4 | 83.9 | 50.1 | 81.7 | 94.2 |
| Phan et al. [7] | 1 EEG + 1 EOG[b] | CNN | 0.779 | 82.3 | 75.5 | **90.6** | 31.9 | **86.8** | 86.7 |
| Yildirim et al. [10] | 1 EEG | CNN | 0.873 | 89.8 | **97.2** | 88.8 | 48.2 | 84.2 | 77.1 |
| Proposed | 1 EEG + 1 EOG | CNN | **0.889** | **91.1** | 96.1 | 81.2 | **71.5** | 75.4 | **95.8** |

a. Electroencephalogram, b. Electrooculogram, c. Convolutional neural networks, d. Bidirectional-long short-term memory.

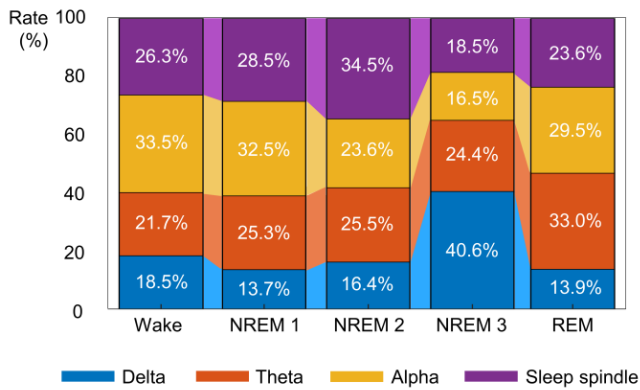

Figure 4. The PSD ratio according to the frequency domain of each class. Each bar represents the ratio occupied by the PSD value of each frequency domain in the corresponding sleep stage.

There are some limitations to this study. First, the proposed method showed relatively low performance at the N2 and REM. It is thought to a trade-off caused by the classifier learning to fit more N1. Second, the classification performance of N1 is still relatively lower compared to other stages. This is because the number of samples in the N1 stage is too small. The imbalance in the class ratio is associated with classification performance [27-28]. It is necessary to classify the sleep stage using the data augmentation to solve the class imbalance. Lastly, we only compared with the studies used sleep-edf dataset, but it needs to compare with several studies using other sleep datasets.

In conclusion, the proposed framework is valuable in that it does not change the whole network, but just extend the kernel size as the added number of bandpass filtered signals by applying the optimal feature. Therefore, our study would apply to other frameworks to improve the performance in sleep stage classification for lessening the burden of sleep experts.